# Gravity influence in one-dimensional blood flow modeling

Oleg Shramko[a]*, Andrey Svitenkov[a], Pavel Zun[a]

[a]*ITMO University, Kronverksky Pr. 49, bldg. A, St. Petersburg 197101, Russia*

**Abstract**

One-dimensional blood flow model accuracy has been verified in many studies. This work is about introducing gravity into a one-dimensional model. For this purpose, gravitational force was introduced into the existing model. The boundary conditions must also be adjusted to account for gravity. For this purpose, a method for calculating arterial resistance of the terminal arteries during gravity changes has been developed and presented in this article. The technique is based on the idea that the human body responds to changes in gravity in such a way as to maintain the volume of blood delivered to the organs. Experiments have shown that not only the pressure changes, but also the flux and pressure waveforms change when the flow is preserved. The method of adjusting the boundary conditions is such that the flow almost does not change when gravity appears, but it is possible to estimate the change in pressure dynamics. This methodology can be used in different cases related to gravity.

## 1. Introduction

In the past 10 years blood flow (BF) modeling became a routine and powerful tool for circulatory system studies for scientific applications. There is also a growing interest in the use of BF modeling in medicine applications. Virtual FFR can be referenced here as the most typical example, and many other applications and studies have been suggested: modeling of aneurysms, blood clot travel, targeted drug delivery, etc.

---

* Corresponding author. Tel.: +7_952-237-2848.
  E-mail address: oashramko@itmo.ru

One-dimensional model of BF holds a special place here since it provides enough computational efficiency to simulate whole body high-detailed arterial system. One-dimensional model accuracy has been verified in many studies and found comparable with uncertainties of model parameters. Sometimes there are attempts to improve one-dimensional model accuracy by using a viscoelastic arterial model [1] advanced bifurcation models [2] or by other methods [3], but in our opinion the major point of interest is currently focused around useful modeling applications of the 1D approach. Populational and sensitivity studies of blood flow patterns [4] are most typical and wide presented, but they provide fundamental rather than practical results.

This paper is aimed to expand 1D BF modeling approach application area by introducing gravity. Possibility of including gravity in 1D model is not novel and is mentioned frequently, however we are not aware about examples of full-body studies where gravity influence would be considered. Such investigation of separate regions of circulatory system (cerebral for example) are presented [5], but gravity influence became negligible at the spatial scale of the arterial model. Gravity consideration could make possible studies of, for example, artificial G-load on circulatory system functioning. Additionally, it generally improves the full-body blood flow mode accuracy under normal conditions.

The possible reason why gravity has not become typical in BF models is probably that the boundary conditions of 1D arterial model should be corrected then as well. Common methodology for boundary parameters estimation does not provide a gravity term and is not applicable in this case [6]. We also understand that the self-regulation system of peripheral vessels changes boundary resistance (and compliance) with the changing of the gravity field. Studies of the gravity influence would be pointless in absence of outlet boundary conditions regulation model that makes the problem more complex than just a modification of the 1D BF governing equations.

In this paper we suggest the idea and methodology for how gravity field can be considered in 1D BF model with the boundary conditions (peripheral vessels) reaction taken into account. We demonstrate our concept on well-known ADAN56 model of human arterial system for verification purpose, implying that a high-detailed arterial model can be used as well. We believe that the suggested idea could be useful as a basic approach for boundary conditions reaction model in future development of the advanced BF models with gravity field consideration.

**Nomenclature**

| | |
|---|---|
| $U$ | Average blood flow velocity, m/s |
| $P$ | Pressure, Pa |
| $A$ | Artery lumen area as a function of P, m$^2$ |
| $A_0$ | Artery lumen area at zero pressure level, m$^2$ |
| $E$ | Young modulus of the arterial wall, Pa |
| $\rho$ | Blood density, kg/m$^3$ |
| $\mu$ | Blood viscosity, Pa·s |
| $h$ | Artery wall thickness, m |
| $R$ | Artery radius, m |
| **r** | 3D space position of the arterial network point, m |
| **g** | Acceleration of gravity, m/s$^2$ |
| $T$ | Pulse time period, s |
| $x$ | Coordinate along the artery, m |
| $t$ | Time, s |

## 2. Methodology

### 2.1. The general idea

Accounting for the gravity influence in blood flow simulations of human circulatory system requires not only inclusion of the gravitational force in the calculations. Additionally, the human body reacts to external changes and adjusts to them. In particular, in order to maintain the same volume of blood flowing to the tissues and organs, the arteries are able to change their resistance [7]. This includes the smallest arteries, arterioles and capillaries, which are considered the terminal arteries for the 1D arterial system. These arteries act as boundary conditions for the 1D model, and can be represented in a 0D form similar to electrical circuits. For more correct accounting of gravity's influence on blood flow, it is also necessary to change the boundary conditions in accordance with changes in gravity, to account for the vessel adaptation.

In the paper [7] authors simulate the influence of microgravity (0G) on the blood flow in the whole human circulatory system. Authors say that arterial resistances were changed according to the body region (+10% for vertebral and carotid arteries; −10% for the lower body), as observed after long exposure to weightlessness. We propose another technique for adjusting the boundary conditions, which is as follows. Starting from the fact that when the gravity changes, the body will attempt to keep the organs supplied with the same volume of blood, we adjust the boundary conditions in such a way that the flow does not change. The body will try to fully compensate for the external influence, and our model implies that the body has unlimited possibilities in this. In reality, the body's capabilities are limited, but we consider that there are no other principles which could be used as a basis for estimation of adjustment of boundary conditions, and study the cases where the body can fully compensate.

*2.2. Model implementation*

The ADAN56 arterial model includes 30 bifurcations and 31 outlets. We initially take all parameters of the model from the benchmark paper [8] to verify the simulation results and make them easily repeatable. Additionally, this provides a better understanding of how the flow patterns change with BC parameters optimization with or without gravity taken into account. Modeling of better detailed arterial networks can be implemented with the same approach.

Typical 1D BF model was used in our simulations. The system of the governing equations was expressed relative to pressure $P$ and average velocity $U$ [8]:

$$\begin{cases} \frac{\partial A}{\partial t} + \frac{\partial (AU)}{\partial x} = 0 \\ \frac{\partial U}{\partial t} + U\frac{\partial U}{\partial x} + \frac{1}{\rho}\frac{\partial P}{\partial x} = \frac{f(U)}{\rho} \end{cases} \quad (1)$$

where the friction term $f(U)$ is defined in the assumption of a Poiseuille parabolic velocity profile $f(U) = -8\mu\pi U$. Relation between the pressure $P$ and the vessel lumen area $A$ is:

$$P(A, x) = P_0 + \frac{4\sqrt{\pi}E(x)h(x)}{3A_0(x)}\left(\sqrt{A} - \sqrt{A_0(x)}\right) + \boldsymbol{r} \cdot \boldsymbol{g} \cdot \rho \quad (2)$$

Here we add an additional gravity term where **r** is a 3D space position of the artery elementary segment and **g** is a vector of acceleration of gravity.

To reduce the number of parameters in the model we use a phenomenological relationship between the reference radius of the vessel and the wall thickness [8]:

$$h = R_0\left[\tilde{a}\exp(\tilde{b}R_0) + \tilde{c}\exp(\tilde{d}R_0)\right] \quad (3)$$

where h is the wall thickness, $R_0$ is the reference radius of the vessel, $\tilde{a} = 0.2802$, $\tilde{b} = -5.053\ cm^{-1}$, $\tilde{c} = 0.1324$, $\tilde{d} = -0.1114\ cm^{-1}$.

Table 1 includes values of the rest of model parameters.

Table 1. Parameters of the model for all verification tests

| Property | Value |
|---|---|
| Blood density, $\rho$, kg·m$^{-3}$ | 1040 |
| Blood viscosity, $\mu$, mPa·s | 4.0 |
| Young's modulus, $E$, kPa | 225.0 |
| Acceleration of gravity, **g**, m/s$^2$ | 0 or 9.8 |
| Space discretization step | 2.5mm |
| Time discretization step | 0.01ms |
| Courant number | <0.4 |

McCormack finite-difference scheme was used for numerical solution of the system (1).

The system of equations for arterial branching includes pressure consistency, flux conservation and continuous propagation of solution characteristics in branches. This approach is also absolutely typical in 1D BF modeling [9].

For arteries outlets 3-parameter lumped BC have been used: peripheral hydrodynamic resistance $R_1$, peripheral vessels compliance $C$ and another hydrodynamic resistance $R_2$ parallel with the compliance. The value of $R_1$ parameter can be defined from assumption of minimized high-frequency reflections in accordance to [6]. The boundary parameter $R_2$ was adjustable.

So, the total terminal resistance of one outlet is equal:

$$R = R_1 + R_2 \tag{4}$$

After the change of gravity the total terminal resistance of one outlet changes as following:

$$R' = R \cdot \frac{(\Delta P_0 - P_{term\,av}) + \Delta P_g}{(\Delta P_0 - P_{term\,av})} \tag{5}$$

where $\Delta P_0$ is the pressure drop between the heart and veins in case with no gravity, $\Delta P_g$ is the pressure of the liquid column, appearing when there is gravity, $P_{term\,av}$ is the average for the cardiac cycle pressure in the corresponding terminal artery.

We calculate $\Delta P_0$ as following:

$$\Delta P_0 = \frac{1}{T} \cdot \frac{\int_0^T P_h Q_h\, dt}{\int_0^T Q_h\, dt} \tag{6}$$

where $P_h$ is the pressure in the heart, $Q_h$ is the flux in the heart, $T$ is the time period (cardiac cycle). In our calculation we use pressure and flux in the first point of aorta as $P_h$ and $Q_h$, respectively.

The pressure of the liquid column is equal to:

$$\Delta P_g = \boldsymbol{r} \cdot \boldsymbol{g} \cdot \rho \tag{7}$$

where the vector $\boldsymbol{r}$ is the radius-vector of the point of interest in arterial topology (relative to the location of the heart), the vector $\boldsymbol{g}$ is the vector of gravitational acceleration.

So, we can find new $R_2$ for every terminal artery and then do the simulation of blood flow.

## 3. Calculation results

For the case study we have chosen 3 model configurations:
1. No gravity, no BC parameters optimization (BC parameters values are taken from the reference [8]);
2. Gravity, no BC parameters optimization (BC parameters values are taken from the reference [8]);
3. Gravity, BC parameters optimization.

The third case provide the model of circulatory system reaction on changing of the gravity field (upright position) while the first two are reference cases. Boundary conditions parameters from the first configuration were input parameters for recalculation of terminal arterial resistances for the third configuration.

For the verification of the simulation implementation of gravity influence on blood flow we checked two things. First, the pressure in legs is greater in case of gravity directed to the legs than one in case of zero gravity. Second, with gravity there is an additional $\rho g h$ term:

$$(P_2' - P_1') - (P_2 - P_1) = \rho g h \tag{8}$$

where $P_1$ is the pressure in the aorta without gravity, $P_2$ is the pressure in the leg without gravity, $P'_1$ is the pressure in the aorta with gravity directed to the legs, $P'_2$ is the pressure in the leg with gravity directed to the legs, h is the height of liquid column.

So, average values for the heart cycle are: $P_1 = 13900\, Pa$, $P_2 = 12450\, Pa$, $P_{1'} = 10400\, Pa$, $P_{2'} = 21050\, Pa$, $\rho g h = 12800\, Pa$, corresponding to a column of blood approximately $1.26\, m$ high. This way, the simulation implementation of gravity influence is verified.

The Figure 1 presents simulation results for the pressure and flux values in upper arteries over one cardiac cycle, after the simulation has stabilized.

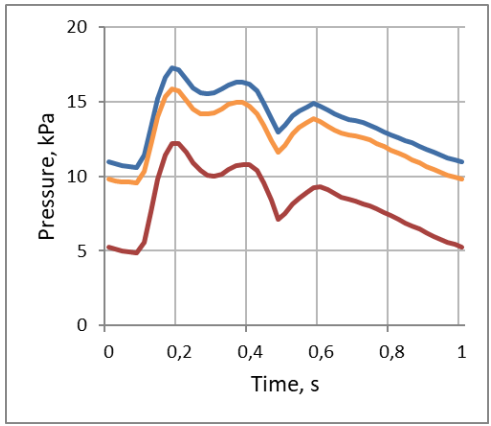
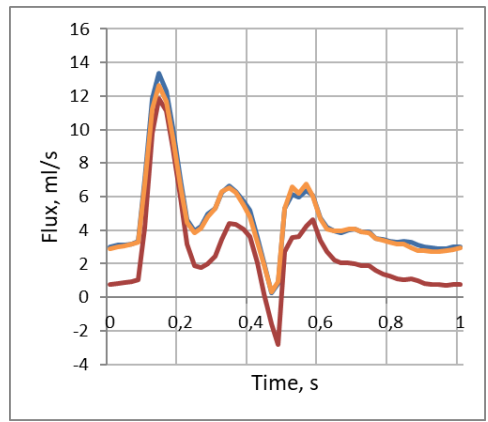

(a) Left Internal Carotid

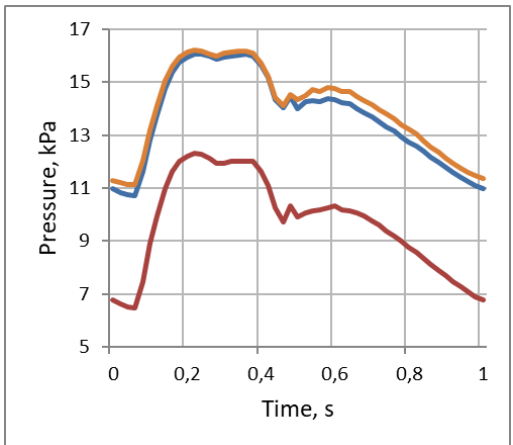
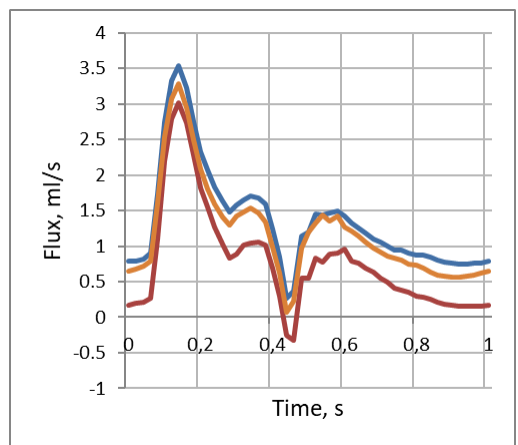

(b) Left Vertebral

Fig. 1. ADAN56 model. Pressure (left) and flow (right) waveforms in the point of two vessels: (a) left internal carotid, (b) left vertebral. Blue color corresponds to the case without optimization and without gravity, red color corresponds to the case without optimization and with gravity, orange color corresponds to the case with optimization and with gravity.

The Figure 2 presents simulation results for the pressure and flux values in lower arteries.

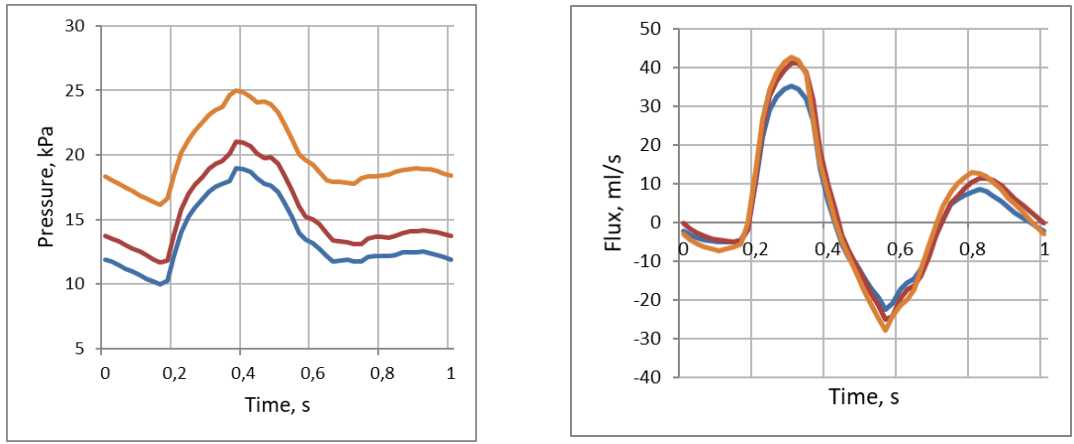

(a) Right Femoral

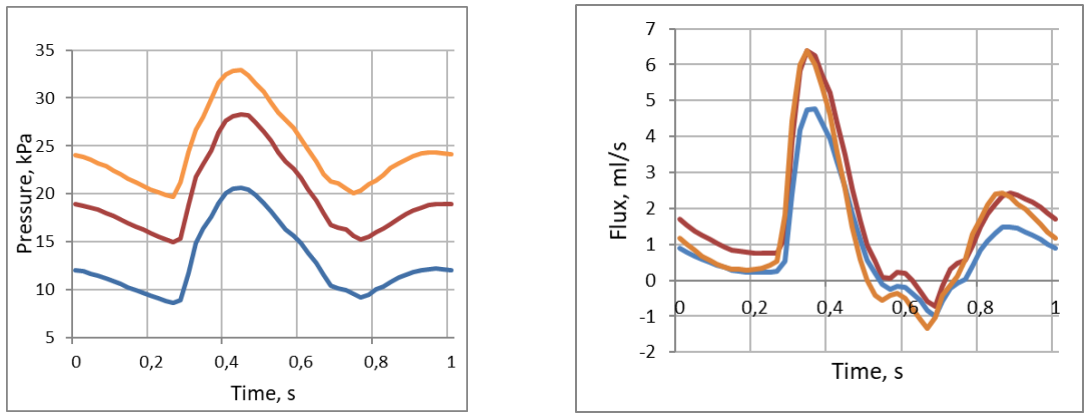

(b) Right Anterior Tibial

Fig. 2. ADAN56 model. Pressure (left) and flow (right) waveforms in the point of two vessels: (a) right femoral, (b) right anterior tibial. Blue color corresponds to the case without optimization and without gravity, red color corresponds to the case without optimization and with gravity, orange color corresponds to the case with optimization and with gravity.

The Figure 3 presents simulation results for the pressure and flux values in middle level arteries.

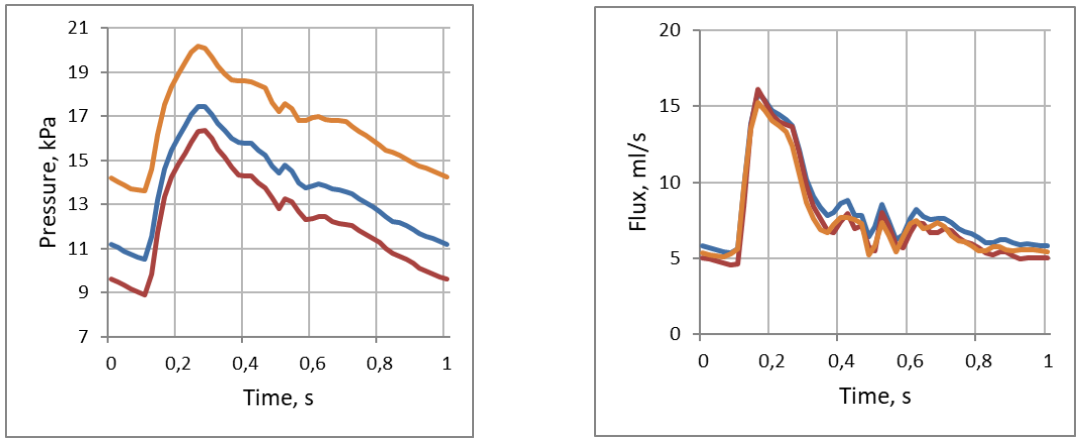

(a) Common Hepatic

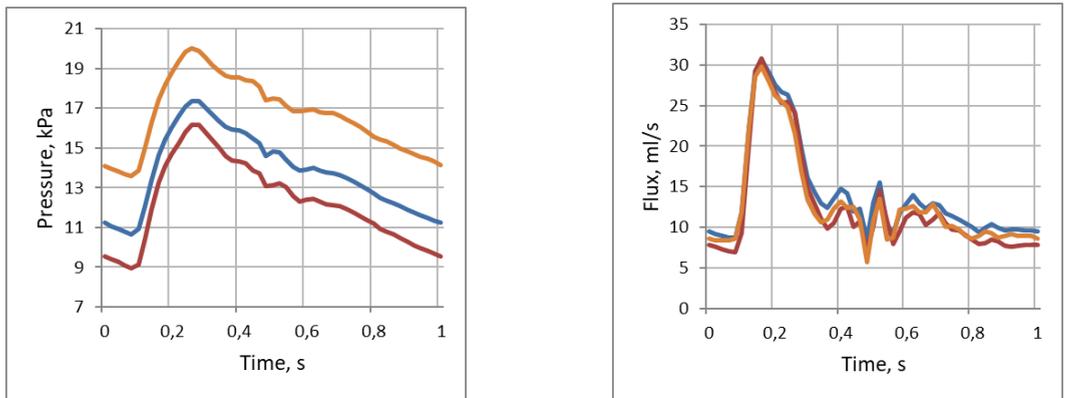

(b) Celiac Trunk

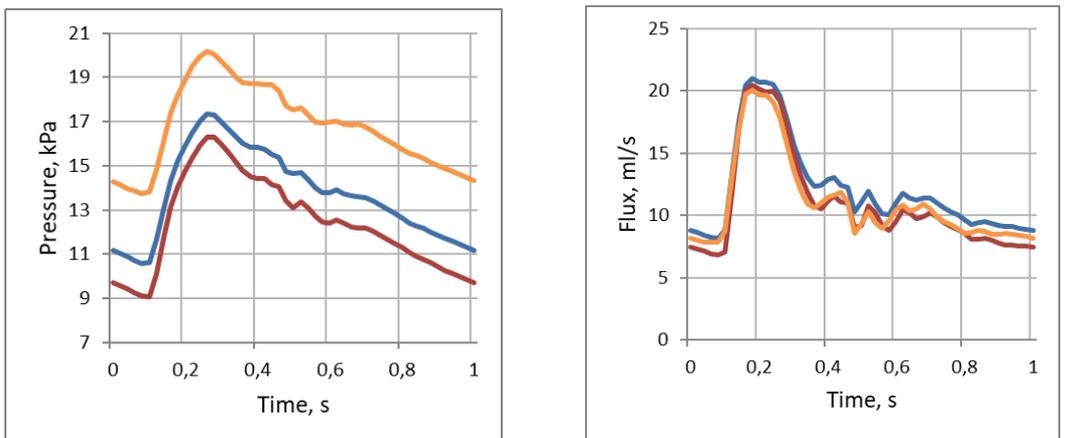

(c) Left Renal

Fig. 3. ADAN56 model. Pressure (left) and flow (right) waveforms in the point of two vessels: (a) common hepatic, (b) celiac trunk, (c) left renal. Blue color corresponds to the case without optimization and without gravity, red color corresponds to the case without optimization and with gravity, orange color corresponds to the case with optimization and with gravity.

The Figure 4 presents simulation results for the pressure and flux values in the hand artery.

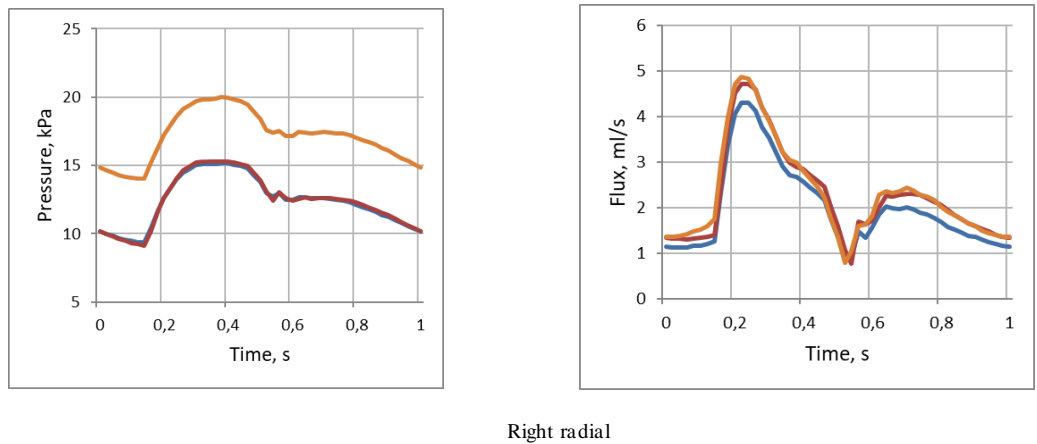

Right radial

Fig. 4. ADAN56 model. Pressure (left) and flow (right) waveforms in the point of right radial artery. Blue color corresponds to the case without optimization and without gravity, red color corresponds to the case without optimization and with gravity, orange color corresponds to the case with optimization and with gravity.

## 4. Discussion

As we see, results show that circulatory system tends to increase the pressure to save the same flux when gravity is introduced in the model.

It seems most important to mention the case of the carotid and vertebral arteries, because they supply blood to the brain. Figure 1 shows that the flux in these arteries is noticeably greater after adjusting the boundary conditions compared to the case with gravity, but without adjusting the boundary conditions. It becomes close to the flux in the case of absence of gravity and with reference values of boundary conditions. At the same time, we can see, as Figure 2 shows, that the flow in the leg arteries as a whole also became closer to the flow corresponding to the first configuration. Similar flow changes, but on a smaller scale, because the difference between the cases of the first and second configurations is much smaller there, can be seen in the rest of the graphs (see Figures 3 and 4). And in all cases, we can see that the adjustment of the boundary conditions results in an increase in pressure. However, not only the pressure changes. There are also some changes in the blood flow waveform. Likely, with a greater gravity force the changes of waveforms will be greater, too.

## 5. Conclusions

We introduced the gravity influence in the blood flow model, verified it for three points of body with simple example based on the addition of liquid column in the calculations. We presented the methodology for simulation the reaction of the body on the changing gravity. This methodology is based on the assumption that human body tends to save the same blood flow when some changes occur, for example change of gravity. Our method of adjusting the boundary conditions is such that the flow almost does not change when gravity appears, but it is possible to estimate the change in pressure dynamics. It allows us to estimate not only how the pressure will change while maintaining the same flow, but also allows you to model how the flow and pressure waveforms will change. This methodology can be used in different cases related to gravity.

The principle underlying the technique (preservation of the volume of blood delivered to the organs) seems to us to be the only one on the basis of which the boundary conditions can be adjusted. In our model, there is no limit to the body's adjustment of terminal resistances. In fact, the possibilities of the organism are limited, and it is the cases when compensation mechanisms fail to manage already that are of the greatest medical interest. However, such studies require clinical medical data.

**Acknowledgements**

This research was supported by The Russian Science Foundation, Agreement # 20-71-10108 (29.07.2020).